\documentclass[aps,prb,twocolumn,superscriptaddress,floatfix,longbibliography,showkeys,showpacs, 10pt]{revtex4-2}
\usepackage{color} 
\usepackage{soul} 
\usepackage{braket}
\usepackage[utf8]{inputenc}
\usepackage{amsmath,amssymb, mathtools}
\usepackage{graphicx}
\usepackage{color}
\usepackage{bm}
\usepackage{textcomp}
\usepackage{tikz}
\usetikzlibrary{scopes}
\usepackage{enumitem}
\usepackage{subfigure}
\usepackage{pgfplots}
\usepgfplotslibrary{fillbetween}
\pgfplotsset{width=9.2cm,compat=newest}
\newcommand{\uvec}[1]

\def\ca{\chi_a(\omega)}
\def\cb{\chi_b(\omega)}
\def\a{\hat{a}}
\def\b{\hat{b}}
\def\ai{\a_\textrm{\tiny{in}}}
\def\bi{\b_\textrm{\tiny{in}}}
\def\Sb{S_{\b\b}(\omega)}

\def\Sai{S_{\ai\ai}(\omega)}
\def\Sbi{S_{\bi\bi}(\omega)}

\def\go{g^{(1)}}
\def\gt{g^{(2)}}
\def\a{\hat{a}}
\def\at{\hat{a}^\dagger}
\def\b{\hat{b}}
\def\bt{\hat{b}^\dagger}
\def\Hi{\hat{H}_I}
\def\tr{\textrm{Tr}}
\def\rdm{\rho_{\textrm{\tiny{DM}}}}
\def\neff{\bar{n}_{\textrm{\tiny{eff}}}}
\def\tr{\textrm{Tr}}
\def\ri{|0\rangle\langle0| \otimes \rho_E}
\def\Am{\varphi^{\scalebox{0.5}{(-)}}}
\def\Ap{\varphi^{\scalebox{0.5}{(+)}}}
\def\ti{t_0}
\def\tf{\ti+\Delta t}
\def\hc{\textrm{h.c.}}
\def\tia{t_a}
\def\tfa{t_a + \Delta t}
\def\tib{t_b}
\def\tfb{t_b + \Delta t}
\def\ba{\hat{b}_a}
\def\bb{\hat{b}_b}
\def\bta{\hat{b}_a^\dagger}
\def\btb{\hat{b}_b^\dagger}
\def\tp{t'}
\def\tdp{t''}
\def\ttp{t^{\scalebox{0.5}{(3)}}}
\def\tqp{t^{\scalebox{0.5}{(4)}}}
\def\ia{I_a}
\def\ib{I_b}
\def\TF{\ti + \Delta T}

\begin{document}

\title{Characterizing the quantum properties of ultralight dark matter - an open quantum systems approach }

\author{Jose-Daniel Bernal}
\thanks{The authors contributed equally to this paper.}
\affiliation{Department of Electrical and Computer Engineering, University of Delaware, Newark, DE 19716, USA}
\affiliation{Departamento de Física, Universidad Nacional de Colombia, 16486 Bogotá, Colombia}

\author{Ryan B. Petery} 
\thanks{The authors contributed equally to this paper.}
\affiliation{Department of Electrical and Computer Engineering, University of Delaware, Newark, DE 19716, USA}

\author{K. J. Joven} 
\affiliation{Department of Electrical and Computer Engineering, University of Delaware, Newark, DE 19716, USA}
\affiliation{Escuela de Ingeniería Eléctrica y Electrónica, Universidad del Valle, 760032 Cali, Colombia}

\author{ Swati Singh}
\email{swatis@udel.edu}
\affiliation{Department of Electrical and Computer Engineering, University of Delaware, Newark, DE 19716, USA}
\affiliation{Department of Physics and Astronomy, University of Delaware, Newark, DE 19716, USA}

\date{\today}

\begin{abstract}
Obtaining insight into the constituents of dark matter and their interactions with normal matter has inspired a wide range of experimental efforts. Several approaches, particularly those involving searches for ultralight bosonic dark matter (UBDM) fields, involve the use of quantum systems or measurements performed at the limits imposed by quantum mechanics. While a classical treatment of UBDM and its detectors is satisfactory, a fully quantum description would assist in developing future detection strategies. Here, we present an open quantum systems approach that accomplishes this while providing intuition into the quantum nature of the detection process itself. Furthermore, we apply the quantum theory of optical coherence to characterize the statistical properties of the UBDM field. Using representative examples, we show that this theoretical treatment has implications in uncovering signatures of the cosmological production mechanism of the UBDM field and its galactic merger history. By adapting tools from quantum optics, this work will help facilitate the creation of novel methods to extract astrophysically relevant information from correlation measurements.

\end{abstract}

\maketitle

Observations across multiple astronomical scales point to the existence of unknown particles, dubbed dark matter (DM), that dominate the mass of galaxies \cite{rubin1970rotation, markevitch2004direct, tyson1998detailed, hinshaw2013nine, aghanim2020planck}. Obtaining insight into the composition of DM and its couplings to standard model (SM) matter has inspired several experimental efforts utilizing new and existing technology to achieve better sensitivity to the weak signals induced by DM \cite{cooley2022report, antypas2022new, adams2022axion}. Many of these experiments leverage quantum systems and measurement techniques to achieve sensitivities limited by quantum fluctuations \cite{chou2023quantum, bass2024quantum}. Experiments of this type are especially relevant to the search for ultralight bosonic dark matter (UBDM), the lowest available mass range of DM candidates that is composed of a high occupation bosonic field \cite{jackson2023search, chou2023quantum}. 

While a classical treatment of UBDM and its detectors is well-justified, future detection strategies would benefit from a quantum treatment of both the field and its interaction with a, perhaps, quantum detector. Additionally, several well-motivated theories suggest that UBDM may itself be a quantum field that might not have undergone decoherence over cosmological times due to its necessarily weak coupling to normal matter \cite{eberhardt2023numerical, marsh2024measuring, cao2023nonequilibrium}. Implications of the quantum nature of the field can only be studied through a quantum description of it. Fortunately, the field of quantum optics already provides a rigorous formalism for characterizing bosonic fields. 

In this paper, we apply the quantum optics formalism to characterize the UBDM field and its interaction with a detector. Using specific examples, we show that the first and second-order UBDM field correlation functions, which contain astrophysically relevant information, can be accessed via a measurement of the detector correlation functions. As is the case for photons, these correlation functions can provide an unambiguous signature of the non-classical nature of the bosonic field \cite{glauber2006nobel}. While we have chosen representative examples for detector (haloscope) and UBDM states (coherent, thermal) to demonstrate the impact of our approach in uncovering astrophysically and cosmologically relevant information from second-order correlation functions, the formalism presented here can be adapted to include any UBDM candidate, their quantum state and detection scheme. 

We start by assuming UBDM to be a multi-mode quantum field. Given its weak coupling to SM matter, one can formulate the interaction between such UBDM and a detector as an open quantum system. Since the detector has much fewer degrees of freedom, and its time evolution can be measured in the lab, we can treat it as the ``system" interacting with a ``bath" of dark matter. By applying our formalism to a well-motivated UBDM search, the cavity haloscope search for the QCD axion \cite{peccei1977constraints,diluzio2020QCDaxionreview}, we show that this specific search can be modeled as a single harmonic oscillator very weakly coupled to a thermal bath (of axions), a well-studied problem in quantum optics. We first describe this interaction in the Lindblad master equation formalism, which gives the axion field's effect on a quantum detector. We then introduce an alternative model in the Heisenberg-Langevin formalism that allows for the bath itself to be probed. In addition, we highlight the promise of the quantum optics approach for uncovering the astrophysical properties of UBDM by showing that the auto-correlation function ($\gt$) of such UBDM can be used to constrain velocity profile deviations from the standard halo model (SHM) for a thermal state or unequivocally demonstrate the coherent nature of the UBDM field.

To treat the detector and UBDM field quantum-mechanically, we will begin by quantizing the UBDM field in analogy to the quantization of the electromagnetic field in quantum optics \cite{gerry2023introductory, scully1997quantum, meystre2021quantum}. To achieve this, we need to choose a quantization volume $V$ larger than anything we consider the field interacting with, and also such that $V^{1/3}$ is larger than the field's wavelength. In order to mimic the boundary conditions for an electric field in a cavity with perfectly conducting walls, we choose $V$ to be bigger than the size of the galaxy. The field operators of any UBDM field can be written as

\begin{equation}
    \hat{\varphi}=c\sum_\mathbf{k,\, s} \sqrt{\frac{\hbar}{2\omega_{\mathbf{k}} V}} \mathbf{e}_{\mathbf{k},\,s} \left( \hat{a}_{\mathbf{k}, \,s}e^{i(\mathbf{k}\cdot\mathbf{r}-\omega_{\mathbf{k}} t)}+\hc \right), \label{eqn:field}
\end{equation}

where $\hat{a}_{\mathbf{k}, \,s}$ is the annihilation operator of mode $\mathbf{k}$ with polarization $s$, $c$ is the speed of light, $V$ is the quantization volume, $\mathbf{e}_{\mathbf{k},\,s}$ is the polarization vector of mode $\mathbf{k}$ with polarization $s$, and $\omega_{\mathbf{k}}$ is the frequency linked to the momentum $\mathbf{k}$ by the dispersion relation $(\hbar \omega_k)^2=(mc^2)^2+(\hbar |\mathbf{k}| c)^2$. The Hamiltonian for the field is then given by
 $ \hat{H}_{\textrm{\tiny{UBDM}}} = \sum_{\mathbf{k},s} \hbar \omega_{\mathbf{k}}\hat{a}^{\dagger}_{\mathbf{k},s}\hat{a}_{\mathbf{k},s}.$
This is a general approach for quantizing a bosonic field and can be applied to any UBDM candidate, cf. ref. \cite{derevianko2018detecting, ioannisian2017axion}. 

In the following, we limit our scope to a single UBDM candidate, the QCD axion of the Peccei-Quinn theory \cite{peccei1977constraints}, thereby dropping the polarization degree of freedom. We take the detector system to be a haloscope, which is a microwave cavity held at cryogenic temperatures with a strong applied magnetic field, as recently reviewed in ref. \cite{sikivie2021invisible}. These systems are commonly used to search for axions, with two prominent examples being ADMX \cite{du2018search} and HAYSTAC \cite{zhong2018results}. The axion's coupling to the magnetic field (shown below in typical particle physics notation in natural units),

\begin{equation}
    \hat{H}_{I}= \int dx^3 g_{a\gamma\gamma} \hat{\varphi} \hat{\mathbf{E}} \cdot   \hat{\mathbf{B}},
    \label{eqn:Hint_particlePhysics}
\end{equation}

generates a coupling between the microwave cavity's electric field and the axion field with coupling constant $g_{a \gamma\gamma}$ \cite{sikivie2021invisible}. After quantizing the electromagnetic field in the microwave cavity, the haloscope detector can be described as a single harmonic oscillator. In the rotating wave approximation, we find the interaction Hamiltonian between the haloscope and the axion field to be 

\begin{equation}
    \hat{H}_{I}= \hbar g\left( \sum_\mathbf{k}\frac{1}{\sqrt{2\omega_\mathbf{k} V}} \hat{a}_\mathbf{k}\hat{b}^\dagger e^{i(\omega_b-\omega_\mathbf{k})t}+ \text{h.c.}\right)  
    \label{eqn:Hint}
\end{equation}

in typical quantum optics notation using SI units, where $g= g_{a\gamma\gamma} (B_0/\mu_0) \sqrt{\hbar \omega_b V'c/\epsilon_0}$, $\omega_b$ is the cavity resonance frequency, $V'$ is the volume of the cavity, $B_0$ is the strength of an applied magnetic field, and $\hat{b}$ is the annihilation operator for the electric field in the microwave cavity. Given the relative strength and directions of the electric and magnetic fields involved in such setups, we have treated the strong applied magnetic field classically in this theoretical treatment. 

\begin{figure}
    \centering
    \includegraphics[width=0.75 \columnwidth] 
    {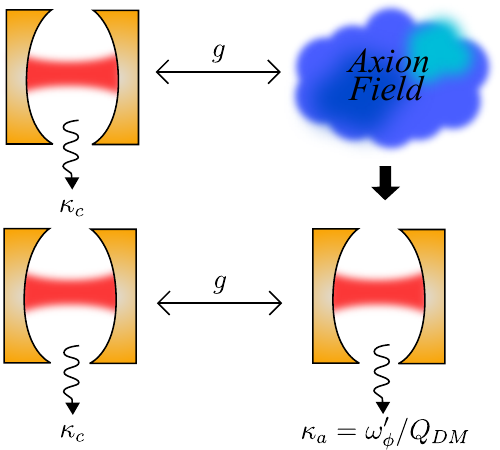}
    \caption{Illustration of the proposed two-cavity model. The multimode axion field is replaced by a second dissipative cavity with similar quality factor.}
    \label{fig:overview}
\end{figure}

The haloscope is very weakly coupled to a system with a much larger number of degrees of freedom such that the Markovian assumption is valid (see Appendix \ref{app:markov}). We also expect that the haloscope's effect on the axion field throughout the interaction to be negligible. We also approximate that the axion field's state will be constant in the interaction picture (known as the Born approximation \cite{breuer2002theory}) and will remain unentangled over the entire interaction, including the initial state. We have then satisfied the requirements to apply an open quantum systems (OQS) formalism \cite{haroche2006exploring}. The haloscope can be considered to be coupled to two separate baths: the typical noisy environment and the axion field. Similar OQS analysis can be applied to most UBDM direct detection experiments, albeit with different interaction Hamiltonians.

We start our OQS analysis by writing down the Lindblad equation for the ideal system and discussing the intuition provided by treating the axion field as a hot thermal bath. However, contrary to the case of typical quantum computing platforms, where coupling to a bath leads to deleterious effects and efforts are made to mitigate it, here we wish to probe properties of the bath itself, which may be in an exotic quantum state. This necessitates the use of Heisenberg-Langevin formalism. After expanding on the limitations of the Lindblad approach, we develop an approximate set of Heisenberg-Langevin equations that capture the essential spectral features of both the detector and the axion bath.

The Lindblad equation describing the evolution of an ideal haloscope interacting with the axion field bath is given by 

\begin{equation}
    \begin{split}
    \dot{\rho}(t) =& -\frac{i}{\hbar}[\hat{H}_c, \rho(t)]\\
    &+ \left(\frac{g}{c}\right)^2\frac{k_b}{4\pi} \left(\b\rho\bt - \frac{1}{2}\left\{ \bt\b,\rho \right\} \right)\left(\neff+1\right)\\
    &+ \left(\frac{g}{c}\right)^2\frac{k_b}{4\pi}\left(\bt\rho\b - \frac{1}{2}\left\{ \b\bt,\rho \right\} \right)\neff,
    \label{eqn:Lindblad}
    \end{split}
\end{equation}

where $\hat{H}_c = \hbar \omega_b \hat{b}^{\dagger}\hat{b}$ is the Hamiltonian of the cavity, and $k_b$ is the momentum related to the cavity frequency $\omega_b$ by the dispersion relation. The derivation of this equation is relegated to Appendix \ref{app:lindbladDerivation}. This is mathematically equivalent to a thermal Lindblad master equation with an effective average particle number 

\begin{equation}
    \neff = \frac{(2\pi)^2\rdm}{2m_ac^2}\int d\Omega\, f(k_b, \theta, \phi),
\end{equation}

where the momentum distribution $f$ is defined by

\begin{equation}
    \langle \at_\mathbf{k}\a_\mathbf{k} \rangle = \frac{(2\pi)^3\rdm}{m_ac^2}f(\mathbf{k}),
\end{equation}

with $\rdm$ being the local UBDM energy density, and $m_a$ being the axion mass. The effective particle number ranges from the order of $10^{92}$ to $10^{4}$ over the entire UBDM mass range ($10^{-22}$ - $1$ eV). Even though converting this effective particle number to a reservoir temperature gives unphysical results, it is instructive to model the detector-axion coupling as a very weak coupling to a very hot bath. Since the effective average particle number of the bath is very large, the classical treatment of the field is well-motivated and valid, both for cosmology \cite{guth2015dark, hertzberg2016quantum} and modeling most UBDM direct detection setups \cite{lehnert2022quantum, beckey2023quantum}. However, as we discuss below, a quantum description of the UBDM interaction with a detector can provide insight into the properties of the expected signal for some experiments.


The beamsplitter interaction in eq. \ref{eqn:Hint} tells us that the haloscope will exchange quanta with all of the axion modes it is coupled to. This quanta exchange is an inherently stochastic process, and the uncertainty in when quanta are exchanged, and between which modes, manifests itself as a stochastic UBDM signal. This effect has been described before in a classical context \cite{centers2021stochastic}, but our Lindblad formalism provides a first principles model for how the state of the detector is affected by the interaction with the axion field, and an intuition into the stochastic nature of the expected UBDM signal on short time scales. As the Markovian approximation is valid for several current UBDM detectors, our Lindblad approach defined above will enable the use of standard OQS numerical methods, widely used in the context of quantum computing, to also model UBDM detection.

In deriving the Lindblad equation, we must consider our haloscope coupled to oscillators at \textit{all} frequencies, effectively giving us a Markovian description of the bath, which translates to a bath of correlation time zero. However, the axion field has a non-zero correlation time and is fairly localized in frequency space with its $Q$ factor of approximately $10^6$ \cite{derevianko2018detecting}. Thus, if we are considering the axion field as a bath, it would be non-Markovian. The Lindblad equation then misses important physical properties of our bath and is insufficient for probing the bath itself. Additionally, the Lindblad master equation derivation assumes that the axion field is in a diagonal density matrix state, however there is reason to believe that may not be the case \cite{boehmer2007can, sikivie2009bose, kopp2022nonclassicality, eberhardt2023numerical}. 

Thus, we seek an alternate approach that will both be applicable for any axion field state, and preserve the non-Markovian nature of the interaction. An open quantum systems approach that satisfies the first condition is the Heisenberg-Langevin formalism \cite{meystre2021quantum}. The equations of motion are here written in the Heisenberg picture, allowing dynamics to be solved using any bath state. However, it fails the second condition, as it requires coupling to oscillators over a bandwidth at least as large as the haloscope bandwidth \cite{scully1997quantum}. Since the quality factors of most haloscopes are lower than the axion field's $Q_a \approx 10^6$, this is not the case. To satisfy the second condition while still working within the Heisenberg-Langevin formalism, we realize that the axion lineshape can be well approximated by a Lorentzian of the same quality factor. We then approximate the axion field as a second cavity with a decay rate equal to $\kappa_a = \omega_\phi'/Q_a$, which is coupled to the haloscope through a beamsplitter interaction (see fig. \ref{fig:overview}). Here, $\omega_\phi'$ is the axion Compton frequency Doppler shifted by the Earth's velocity in the galactic reference frame as defined in \cite{derevianko2018detecting}. This is similar to the cascaded systems approach utilized in quantum noise studies \cite{gardiner2004quantum}, where one system is used to ``create" the input field for another system. Here, we are using a second dissipative cavity to create a field with similar properties to the axion field, and then coupling it to the haloscope field. Assuming the haloscope has its own dissipation rate $\kappa_c$, we can write the equations of motion for this system,

\begin{equation}
    \begin{split}
        \dot{\b} &= -i\omega_b\b -ig\a - \frac{\kappa_c}{2}\b + \sqrt{\kappa_c}\bi,\\ 
        \dot{\a} &= -i\omega_\phi'\a -ig\b -\frac{\kappa_a}{2}\a + \sqrt{\kappa_a}\ai.
        \label{eqn:twoCavity}
    \end{split}
\end{equation}

Note that this $\a$ operator is unrelated to the $\hat{a}_\mathbf{k}$ from earlier, but is instead the annihilation operator of the cavity representing the axion field for the purposes of this model. As this is written in the Heisenberg-Langevin formalism, it remains general in the bath state. Additionally, it approximately preserves the non-Markovian nature of the interaction. Thus, we have satisfied our original two conditions and derived a model that describes the system's dynamics in a simple set of equations. 

In the small coupling limit and under the Born approximation, we find 

\begin{equation}
\begin{split}
    \Sb = \left|\cb\right|^2\left[\kappa_c\Sbi + g^2\left|\ca\right|^2\Sai\right],
    \label{eqn:correlationLink}
\end{split}
\end{equation}

where

\begin{equation}
    \chi_{b(a)}(\omega) = \frac{1}{i(\omega_{b(\phi)}-\omega) + \frac{\kappa_{c(a)}}{2}},
\end{equation}

and $\Sb$ and $\Sai$ are the unnormalized haloscope and axion cavity input field first-order correlation functions, respectively, in the frequency domain. As seen in eq. \ref{eqn:correlationLink}, the correlation functions of the haloscope are related to the axion correlations, i.e. we can measure the axion correlations by measuring the haloscope correlations. 

Experimental access to the axion field correlation functions enables us to characterize the quantum statistical properties of the UBDM field itself by applying tools from the quantum theory of optical coherence. Originally developed to characterize the light from a laser and understand correlations in thermal light, Glauber's landmark work has improved our understanding of the quantum nature of light \cite{glauber1963quantum}. It has since been applied to demonstrate coherence in light \cite{arecchi1966time}, cold atoms \cite{ketterle1997coherence, ottl2005correlations, schellekens2005hanbury, jeltes2007comparison}, and mechanical systems \cite{cohen2015phonon} among others, while finding new applications in nuclear physics \cite{baym1998physics} and fluid dynamics \cite{berne2000dynamic}. Here we apply his theory to characterize the quantum state of an UBDM field. We will focus on $\go(\tau)$ and $\gt(\tau)$, which are defined in terms of the forward and backward evolving parts of the field,

\begin{equation}
\begin{split}
    &\Ap(t) = c\sum_\mathbf{k,\, s} \sqrt{\frac{\hbar}{2\omega_{\mathbf{k}} V}}  \hat{a}_{\mathbf{k}, \,s}e^{i(\mathbf{k}\cdot\mathbf{r}-\omega_{\mathbf{k}} t)},\\
    &\Am(t) = c\sum_\mathbf{k,\, s} \sqrt{\frac{\hbar}{2\omega_{\mathbf{k}} V}}  \hat{a}^{\dagger}_{\mathbf{k}, \,s}e^{-i(\mathbf{k}\cdot\mathbf{r}-\omega_{\mathbf{k}} t)}. \label{eqn:fieldParts}
\end{split}
\end{equation}

The $\go$ and $\gt$ functions for the bosonic field $\hat{\varphi}$ are

\begin{equation}
\begin{split}
    &\go(\tau) = \frac{\langle\Am(t)\Ap(t+\tau)\rangle}{\langle\Am(t)\Ap(t)\rangle},\\
    &\gt(\tau) = \frac{\langle\Am(t)\Am(t+\tau)\Ap(t+\tau)\Ap(t)\rangle}{\langle\Am(t)\Ap(t)\rangle^2}.
\end{split}
\end{equation}

These functions give us information on the quantum state of the field. For example, if the initial production mechanism was the misalignment mechanism, it is possible that the axion field was created in a coherent state \cite{abbott1983cosmological, preskill1983cosmology}, and given its weak coupling to SM matter, it may not have decohered fully. A coherent state will have $|\go(\tau)| = \gt(\tau) = 1$. On the other hand, if the axion field has thermalized, it will be in a multimode thermal state with $|\go(0)| = 1$, which decays to zero as the time lag increases, and $\gt(\tau) = 1 + |\go(\tau)|^2$ for thermal states \cite{loudon2000quantum}. Note that the coherent state has $\gt(0) = 1$, and the thermal state has $\gt(0) = 2$, an effect known as bunching \cite{brown1956correlation}. This allows us to differentiate coherent and thermal (or other diagonal density matrix) states. 

\begin{figure}
    \centering
    \includegraphics[width=1 \columnwidth] 
    {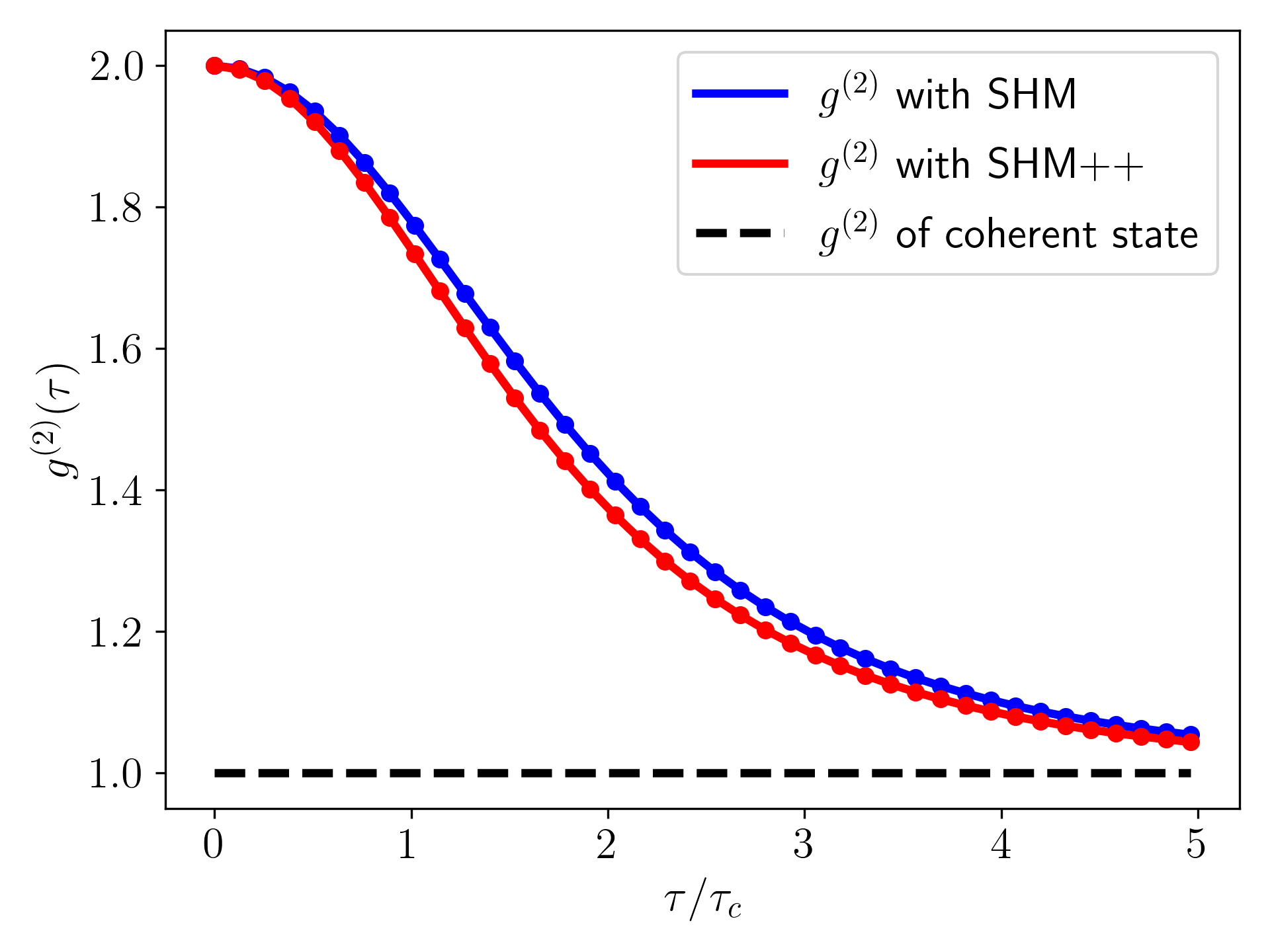}
    \caption{Plot of $\gt$ correlation functions for different axion states. The blue curve is the $\gt$ of a thermal state with a velocity given by the SHM. The red curve is the same, but using the alternative SHM++ \cite{evans2019refinement}, which takes into account the merger history of the milky way. The black curve is the $\gt$ of a multimode coherent state.} 
    \label{fig:g2}
\end{figure}

Even if the axion field is in a completely thermalized state, we now show that the correlation functions can give us astrophysically relevant information. For a thermalized axion field with a velocity profile $f(\mathbf{k})$, the normalized correlations are

\begin{equation}
    \go(\tau) =\left\langle \frac{1}{\omega_k}\right\rangle_f^{-1} \int d^3\mathbf{k} \frac{ f(\mathbf{k}) }{ \omega_{k}} e^{-i\omega_{k}\tau},
\end{equation}

\begin{equation}
    \gt(\tau) = 1 + \left\langle \frac{1}{\omega_k}\right\rangle_f^{-2}  \left | \int d^3\mathbf{k} \frac{f(\mathbf{k})}{\omega_k}e^{-i\omega_{k}\tau} \right |^2.
\end{equation}

Since these expressions depend on the specific velocity distribution, the shapes of these functions can be used to distinguish velocity profiles as seen in fig. \ref{fig:g2}. The blue curve shows the $\gt$ for a thermal axion field assuming the SHM, which assumes a Maxwell-Boltzmann distributed UBDM halo around the galaxy \cite{drukier1986detecting}. This gives the velocity distribution.

\begin{equation}
    f_{\textrm{\tiny{SHM}}}(\mathbf{v}) \propto \exp\left(\frac{-|\mathbf{v}-\mathbf{v}_g|^2}{2v_v^2}\right)\Theta(v_\textrm{\tiny{esc}}-|\mathbf{v}|),
    \label{eqn:SHM}
\end{equation}


where $\mathbf{v}_g$ is earth's velocity in the galactic reference frame, $v_\textrm{\tiny{esc}}$ is the galactic escape velocity, and $v_v = 10^{-3}c$ is the virialization velocity. 
The red curve in fig. \ref{fig:g2} shows the $\gt$ for a thermal axion field assuming the SHM++ (an alternate model taking into account the merger history of the Milky Way) \cite{evans2019refinement}, and the black curve is the $\gt$ of an axion field in a pure coherent state. As demonstrated in fig. \ref{fig:g2}, different velocity distributions give rise to slightly different bunching statistics if the axion field is purely classical, and can confirm the field's coherence.

Measuring these correlation functions require photon counting/intensity measurements, where the probability of detecting a photon in the cavity twice in two disjoint short time periods is proportional to the second-order correlation function, as discussed in detail in Appendix \ref{photonCounting}. In the context of haloscope detectors, intensity measurements could be performed by looking at the loss port \cite{goryachev2019axion}. Two detectors will be required to measure the correlation functions plotted in fig. \ref{fig:g2}. Using interferometry to probe certain properties of the UBDM field has been discussed previously \cite{marsh2024measuring, masia2023intensity, foster2021dark}. However, our approach establishes a rigorous formalism to distinguish between various quantum states of the UBDM field using intensity interferometry.  

From a quantum optics standpoint, both coherent and thermal states discussed above are relatively classical in nature. An unambiguous signature of a non-classical field is $\gt<1$, an effect known as anti-bunching \cite{walls1983squeezed, davidovich1996sub}. While typically observed in low-photon number Fock states, it can also be observed in certain squeezed states. Interestingly, non-linear gravitational (or other) interactions might provide a pathway for a large average occupation number coherent state to evolve into a highly squeezed state \cite{kuss2021squeezing, kopp2022nonclassicality}, particularly given the cosmic time-scales and weak decoherence channels in effect for UBDM fields \cite{cao2022brownian, cao2023nonequilibrium, eberhardt2022quantum, eberhardt2024classical}. 
Analysis of such squeezed states and extracting cosmologically relevant information from $\gt$ functions is the subject of future work.

To summarize, we claim that the UBDM detection problem is well described using the open quantum systems approach, where the highly developed formalism allows for a quantum description of the field's influence on detectors, and focus on the case of QCD axion detection with haloscopes as an illustrative example. The axion field is typically considered as a classical field primarily due to its high occupation number along with the expectation of small quantum corrections \cite{guth2015dark, hertzberg2016quantum}. However, as first demonstrated by Planck, the quantum nature of a bosonic field can also be revealed in one with a large average occupation number \cite{max1901uber, glauber2006nobel}. Other representative examples from optics include spontaneous emission, Lamb Shift, and the Casimir effect \cite{muthukrishnan2017concept}. Similar connections are being made to reveal the quantum nature of gravity \cite{tobar2023detecting, carney2024graviton}, where large occupation number gravitational waves motivated a classical description \cite{caves1980measurement}.

We see this work as a natural and necessary extension to the axion detection problem which allows experimenters to design experiments to efficiently probe more properties of dark matter, such as the field's (possibly non-classical) state and even the merger history of our galaxy. We have also introduced a simple and intuitive model for the axion-haloscope interaction in the Heisenberg-Langevin formalism, which takes into account the non-Markovian nature of the interaction while remaining valid for all axion field states. This quantum description of the field could possibly lead to improved waveform estimation beyond the limits derived in the semiclassical regime \cite{gardner2024stochastic, zhuang2023ultimate}.

Finally, our work, along with other efforts to describe particle physics experiments in quantum optics formalism are adding to the mutually symbiotic relationship between astronomy and quantum optics. From explaining the solar blackbody spectrum \cite{max1901uber} to using auto-correlation measurements to characterize stars \cite{brown1957interferometry, brown1958interferometry, glauber1963coherent}, several critical advances in quantum optics were motivated by astronomical measurements. More recently, efforts to understand the nature of quantum measurements were inspired by gravitational wave detection \cite{caves1980measurement,caves1981quantum, danilishin2012quantum}. By applying the rigorous theoretical formalism developed to characterize the quantum nature of light, our work contributes towards improving our understanding of ultralight dark matter and investigating ways to detect it.

\emph{Acknowledgements:} We acknowledge insightful discussions with  J. Betz and H. Vinck, P. Zoller, P. Meystre, K. Mølmer, G. Rybka, M. Tobar, K. Shultz, D. Budker and A. Derevianko. This work is supported by the National Science Foundation grants PHY-1912480 and PHY-2047707, and the AFOSR DEPSCoR-RC program A9550-22-1-0323.

\bibliography{main}

\appendix

\section{UBDM Field Quantization}

We focus here on the case of a spin-zero field, noting that higher integer spins will result in an extra polarization degree of freedom. The Klein-Gordon Lagrangian density in SI units is given by 

\begin{equation}
    \mathcal{L} = \frac{1}{2}\partial^\mu \varphi \partial_\mu \varphi - \frac{1}{2}\left(\frac{mc}{\hbar}\right)^2\varphi^2.
\end{equation}

From this, we conclude that the units of the field must be $[\varphi] = \sqrt{E/L}$, and add a correction factor of $c\sqrt{\hbar}$ to the typical particle physics quantized field in natural units \cite{srednicki2007quantum}, giving us the final expression in eq. \ref{eqn:field}, keeping in mind the added polarization degree of freedom for vector candidates.

\section{Interaction Hamiltonian}

To derive the interaction Hamiltonian, we start with the axion field's coupling to the electromagnetic field from the Peccei-Quinn theory. Converting eq. \ref{eqn:Hint_particlePhysics} of the main text to SI units, we get \cite{sikivie2021invisible}

\begin{equation}
    \hat{H}_{I}= \sqrt{\hbar c}\int dx^3 g_{a\gamma\gamma} \hat{\varphi} \frac{\hat{\mathbf{E}} \cdot   \hat{\mathbf{B}}}{c \mu_0}.
\end{equation}

We consider the field's interaction with a haloscope, which consists of a cavity of volume $V'$ made of a conductor material upon which a very strong magnetic field is applied. In the following, we will consider only the interaction of the axion field with the fundamental EM mode of a cubic cavity, whose field is written in terms of the creation and annihilation operators $b$ and $b^\dagger$ as

\begin{equation}
\hat{\mathbf{E}}=\left( \frac{\hbar \omega_b}{\epsilon_0 V'} \right)^{1/2}(\hat{b}+\hat{b}^\dagger) \sin(k_Ex) \mathbf{z}, \\
\end{equation}

where $k_E = \pi/L'$ with $L'^3 = V'$. The applied magnetic field is given by

\begin{equation}
\hat{\mathbf{B}}=B_0\mathbf{z}.
\end{equation}

The dot product of the cavity's electric and magnetic field will be zero, so this term in the total magnetic field is neglected. The axion field quantized in a volume $V$ is given by

\begin{equation}
    \hat{\varphi}= c\sum_\mathbf{k}  \sqrt{\frac{\hbar}{2V\omega_\mathbf{k}}}  \left\{ \hat{a}_\mathbf{k}e^{i(\mathbf{k}\cdot\mathbf{x}-\omega_\mathbf{k} t)}+\hat{a}_\mathbf{k} ^\dagger e^{-i(\mathbf{k}\cdot \mathbf{x}-\omega t)} \right\},
\end{equation}

with dispersion relation $(\hbar\omega)^2=(mc^2)^2+(\hbar k c)^2$. We note that the axion field is only occupied in a small range of modes near the axion's Compton frequency. We assume here that our haloscope is resonant at this frequency, $\omega_{\mathbf{k}} \approx \omega_b$. In appendix C, we see $|\hat{H}_I|\ll\hbar/\tau_c$, which implies that the coupling frequency is much less than the Compton frequency and therefore, we can perform the rotating wave approximation and neglect the co-rotating terms \cite{mandel1995optical}. This leads to the interaction Hamiltonian

\begin{equation}
    \hat{H}_{I}= \hbar g\left\{ \sum_\mathbf{k}\frac{1}{\sqrt{2V\omega_\mathbf{k}}} \hat{a}_\mathbf{k}\hat{b}^\dagger e^{i(\omega_b-\omega_\mathbf{k})t}+ \text{h.c.}\right\},  \label{eqn:Hinteraction}
\end{equation}

with $g= g_{a\gamma\gamma} (B_0/\mu_0) \sqrt{\hbar \omega_b V'c/\epsilon_0}$. We see that this is a typical beamsplitter interaction, and the cavity will exchange quanta with the axion field. This is equivalent to eq. \ref{eqn:Hint} in the main text.

\section{Lindblad Markovian Approximation Validity}
\label{app:markov}

For the Lindblad approach to be valid, the differential equation must be discretized with a time step small enough to capture system dynamics and large enough to ignore the bath's memory time. We require that our typical system dynamics timescale $T_r$ be much larger than the axion field's coherence time $\tau_c$. In natural units, our system's timescale can be written as $T_r = 1/|\hat{H}_I|^2\tau_c$, where $|\hat{H}_I|$ is the size of the interaction Hamiltonian matrix elements \cite{haroche2006exploring} (we adopt the notation $|\cdot|$ for the typical magnitude of an operator's matrix elements). The final condition that needs to be satisfied is

\begin{equation}
    \tau_c \ll \frac{1}{|\hat{H}_I|^2\tau_c}.
\end{equation}

Using eq. \ref{eqn:Hint_particlePhysics}, we can bound $|\hat{H}_I|$. The magnitude of the magnetic field matrix elements is $|\hat{B}| = B_0$. Since the haloscope is expected to have a very low occupation throughout an experiment, we estimate $|\hat{E}| = \sqrt{\omega_b/V'}\sin{(k_E x)}$. Since we assume $\omega_b \approx \omega_\phi'$, and in natural units $\omega_\phi' = m_a$ where $m_a$ is the mass of an individual axion, we can write $|\hat{E}| \approx \sqrt{m_a/V'}\sin{(k_E x)}$. We now turn to the magnitude of the axion field matrix elements. We can bound it from above using the typical dark matter number density $\rdm = 0.3\, \mathrm{GeV}/\mathrm{cm}^3$ \cite{catena2010novel}. We expect 

\begin{equation}
    |\hat{\varphi}| \ll \sqrt{\frac{n}{m_a V}},
\end{equation}

where $n = \rdm V/m_a$. We stress that this is an overestimation due to ignoring the mixtures of phases on each term of the sum of eq. \ref{eqn:field}. This gives an upper bound for the magnitude of the interaction Hamiltonian matrix elements,

\begin{equation}
    |\hat{H}_I| \ll 2g_{a\gamma\gamma}\frac{\sqrt{V'n/V}B_0}{\pi}.
\end{equation}

Using the fact that the axion mass is related to the symmetry-breaking energy scale $f_a$ \cite{sikivie2021invisible} and that $|\hat{H}_I|$ is overestimated, we can turn this into a condition on $f_a$. Using values for $V'$ and $B_0$ from HAYSTAC and ADMX, we arrive at the conditions below.

\begin{center}
\begin{tabular}{ c c }
 HAYSTAC: & $f_a < 4 \times 10^{14}$ GeV  \\ 
 ADMX: & $f_a < 1 \times 10^{14}$ GeV    
\end{tabular}
\end{center}

As the HAYSTAC and ADMX experiments are searching for axions in the $f_a \approx 10^{11}$ GeV \cite{backes2021quantum} and $f_a \approx 10^{12}$ GeV \cite{du2018search} regimes respectively, the Markovian assumption is valid for both experiments.

\section{Lindblad Derivation} \label{app:lindbladDerivation}

This treatment is adapted from similar work in ref. \cite{breuer2002theory}. We begin with the interaction Hamiltonian between a cavity (Here a single mode harmonic oscillator) and the axion field (a multimode bosonic field) assuming the axion center frequency is approximately equal to the cavity frequency.

\begin{equation}
    \hat{H}_I = \hbar g \sum_{k} \frac{1}{\sqrt{2V\omega_\mathbf{k}}} \bigg[ \a_k \bt e^{i(\omega_b-\omega_k)t} + \at_k \b e^{-i(\omega_b-\omega_k)t} \bigg].
\end{equation}

As in \cite{breuer2002theory}, we express the equation of evolution for the density matrix (Von Neumann equation) in integrodifferential form up to second-order, perform the Born-Markov approximation (see appendix \ref{app:markov}), and trace over the degrees of freedom of the axion field to get the Redfield equation in the interaction picture

\begin{equation}
\begin{split}
    \dot{\rho} &= -\frac{1}{\hbar^2}\int_{0}^{\infty}ds\,\tr_E\left[ \Hi(t), \left[ \Hi(t-s) , \rho(t)\otimes\rho_E \right] \right].
\end{split}
\end{equation}

This approximation is valid assuming a Markovian bath with small coupling. The commutators will give us a total of four terms. Each of these terms will have some cancellation. Any terms with non-crossed $\a$ terms will go to zero, and all terms with a mix of $\a$ of different modes will cancel assuming a diagonal density matrix. We can cancel those same terms in the case of a general density matrix using the rotating wave approximation.  

The first term is

\begin{equation}
\begin{split}
    -\frac{1}{\hbar^2}&\int_{0}^{\infty}ds\,\tr_E \left[ \Hi(t)\Hi(t-s)\rho(t)\otimes\rho_E \right]\\
    = & g^2\int_{0}^{\infty}ds\, \sum_\mathbf{k} \frac{1}{2V\omega_\mathbf{k}} \bigg{[} \bt\b\rho(t)e^{i(\omega_b-\omega_\mathbf{k})s}\tr\left[ \a_\mathbf{k}\at_\mathbf{k} \rho_E\right]\\ 
    &+ \b\bt\rho(t)e^{i(\omega_b-\omega_\mathbf{k})s}\tr\left[ \at_\mathbf{k}\a_\mathbf{k} \rho_E\right]\bigg{]}.
    \label{eq:lindbladTerm1}
\end{split}
\end{equation}

We then use the identity

\begin{equation}
    \Re\left({\int_0^\infty e^{inx}\,dx}\right) = \pi \delta(n).
\end{equation}

The imaginary part generates a negligible shift in the cavity's energy levels, in analog to a Lamb shift. The RHS of eq. \ref{eq:lindbladTerm1} turns into

\begin{equation}
\begin{split}
    &-g^2\pi \sum_\mathbf{k} \frac{1}{2V\omega_\mathbf{k}} \bigg{[} \bt\b\rho(t) \delta(\omega_b-\omega_\mathbf{k}) \tr\left[ \a_\mathbf{k}\at_\mathbf{k} \rho_E\right]\\
    &+ \b\bt\rho(t) \delta(\omega_b-\omega_\mathbf{k}) \tr\left[ \at_\mathbf{k}\a_\mathbf{k} \rho_E\right]\bigg{]}.
\end{split}
\end{equation}

We now use a property of the Dirac delta function:

\begin{equation}
\begin{split}
    &\textrm{For continuous smooth function $g(x)$}\\
    &\textrm{such that $g(x_0)=0$;}\\
    &\delta(g(x)) = \frac{\delta(x-x_0)}{|g'(x_0)|},
\end{split}
\end{equation}

and rewrite the RHS as

\begin{equation}
\begin{split}
    &-g^2\pi \sum_\mathbf{k} \frac{\omega_b}{2V\omega_\mathbf{k}k_bc^2} \bigg{[} \bt\b\rho(t) \delta(k_b-|\mathbf{k}|) \tr\left[ \a_\mathbf{k}\at_\mathbf{k} \rho_E\right]\\
    &+ \b\bt\rho(t) \delta(k_b-|\mathbf{k}|) \tr\left[ \at_\mathbf{k}\a_\mathbf{k} \rho_E\right]  \bigg{]},
\end{split}
\end{equation}

where $k_b^2 = (\omega_b/c)^2-(m_ac/\hbar)^2$.

We then make the continuous mode approximation 

\begin{equation}
    \sum_\mathbf{k} \to \frac{V}{(2\pi)^3}\int\,d^3\mathbf{k},
\end{equation}

and make a switch to spherical coordinates. 

\begin{equation}
\begin{split}
    &-\frac{g^2\pi}{2(2\pi)^3} \int dkd\theta d\phi\, k^2\sin(\phi)\\
    &\times\frac{\omega_b}{\omega_{k} k_bc^2}\bigg{[}\bt\b\rho(t) \delta(k_b-k)\tr\left[ \a_k\at_k \rho_E\right]\\
    &+ \b\bt\rho(t) \delta(k_b-k) \tr\left[ \at_k\a_k \rho_E\right]\bigg{]}.
\end{split}
\end{equation}

We continue by making the substitutions

\begin{equation}
\begin{split}
    &\tr\left[ \at_\mathbf{k}\a_\mathbf{k} \rho_E\right] = \frac{(2\pi)^3\rdm}{m_ac^2}f(\mathbf{k}),\\
    &\tr\left[ \a_\mathbf{k}\at_\mathbf{k} \rho_E\right] = \frac{(2\pi)^3\rdm}{m_ac^2}f(\mathbf{k}) + 1,
\end{split}
\end{equation}

and take the integral over $k$ to get

\begin{equation}
\begin{split}
    &-\frac{g^2\pi k_b}{2(2\pi)^3c^2} \int d\Omega \bigg{[}\bt\b\rho(t) \bigg{(}\frac{(2\pi)^3\rdm}{m_ac^2}f(k_b, \theta, \phi) + 1\bigg{)}\\
    &+ \b\bt\rho(t) \bigg{(}\frac{(2\pi)^3\rdm}{m_ac^2}f(k_b, \theta, \phi)\bigg{)} \bigg{]}.
\end{split}
\end{equation}

We now define 

\begin{equation}
    \neff = \frac{(2\pi)^2\rdm}{2m_ac^2}\int d\Omega\, f(k_b, \theta, \phi)
\end{equation}

to simplify term 1 to

\begin{equation}
    -\left(\frac{g}{c}\right)^2\frac{k_b}{4\pi}\left[ \bt\b\rho(t)(\neff+1) + \b\bt\rho(t)\neff \right].
\end{equation}

Term 2 of the expression is

\begin{equation}
    \frac{1}{\hbar^2}\int_{0}^{\infty}ds\,\tr_E \left[ \Hi(t)\rho(t)\otimes\rho_E\Hi(t-s) \right].
\end{equation}

Using the same procedure as term 1, we find that term 2 is equal to

\begin{equation}
    \left(\frac{g}{c}\right)^2\frac{k_b}{4\pi}\left[ \b\rho(t)\bt(\neff+1) + \bt\rho(t)\b\neff \right].
\end{equation}

Term 3 is

\begin{equation}
    \frac{1}{\hbar^2}\int_{0}^{\infty}ds\,\tr_E \left[ \Hi(t-s)\rho(t)\otimes\rho_E\Hi(t) \right],
\end{equation}

which evaluates to the same as term 2. 

Term 4 is 

\begin{equation}
    -\frac{1}{\hbar^2}\int_{0}^{\infty}ds\,\tr_E \left[ \rho(t)\otimes\rho_E\Hi(t-s)\Hi(t) \right],
\end{equation}

which evaluates to

\begin{equation}
    -\left(\frac{g}{c}\right)^2\frac{k_b}{4\pi}\left[ \rho(t)\bt\b(\neff+1) + \rho(t)\b\bt\neff \right].
\end{equation}

Putting all terms together and switching out of the interaction picture, we find

\begin{equation}
\begin{split}
    \dot{\rho}(t) = &-\frac{i}{\hbar}[\hat{H}_c, \rho(t)] \\
    &+ \left(\frac{g}{c}\right)^2\frac{k_b}{4\pi}\bigg[ \left(\b\rho\bt - \frac{1}{2}\left\{ \bt\b,\rho \right\} \right)\left(\neff+1\right)\\
    &+ \left(\bt\rho\b - \frac{1}{2}\left\{ \b\bt,\rho \right\} \right)\neff\bigg],
\end{split}
\end{equation}

where $\hat{H}_c = \hbar \omega_b \hat{b}^{\dagger}\hat{b}$ is the Hamiltonian of the cavity, and the Hamiltonian $\hat{H}_{\textrm{\tiny{UBDM}}}$ is defined in the main text. This is mathematically equivalent to a thermal Lindblad master equation with rate $(g/c)^2k_b/(4\pi)$ and particle number $\neff$. This equation is equivalent to eq. \ref{eqn:Lindblad}.

\section{Two-Cavity Model Derivation}

Axions and other UBDM can be thought of as a multimode field, with modes around the Doppler shifted Compton frequency ($\omega_\phi'$) at a width of approximately $\omega_\phi'/Q_{\textrm{DM}}$, where $Q_{\textrm{DM}} \approx 10^6$ is the quality factor of the DM field \cite{derevianko2018detecting}. The axion field is mathematically represented as a collection of harmonic oscillators over some frequency range. We can then think of the axion field as a cavity with resonance frequency $\omega_\phi'$ and a decay rate of $\kappa_a = \omega_\phi'/Q_{\textrm{DM}}$. The dissipation splits the single mode of the cavity into a range of modes around resonance at a width $\kappa_a$, thus giving us another system with a similarly narrow band spectrum to the axion field.

We then consider a haloscope cavity (such as ADMX) coupled to the axion field through a beamsplitter interaction with coupling $g$

\begin{equation}
    \Hi = \hbar g[\b^\dagger\a + \a^\dagger\b],
\end{equation}

where $\a$ and $\b$ are the ladder operators of the axion field and haloscope cavity, respectively. We also assume the haloscope cavity has a dissipation rate $\kappa_c$.

The model outlined above gives us a pair of coupled differential equations for $\a$ and $\b$ in the Heisenberg-Langevin formalism \cite{meystre2021quantum},

\begin{equation}
    \dot{\b} = -i\omega_b\b -ig\a - \frac{\kappa_c}{2}\b + \sqrt{\kappa_c}\bi,
\end{equation}
\begin{equation}
    \dot{\a} = -i\omega_\phi'\a -ig\b -\frac{\kappa_a}{2}\a + \sqrt{\kappa_a}\ai.
\end{equation}

The equation for $\b$ can be solved in the Fourier domain:

\begin{equation}
\begin{split}
    \b =& \,\sqrt{\kappa_c}\chi_b(\omega) \bi\\
    &+ - i g \sqrt{\kappa_a} \chi_b(\omega) \chi_a(\omega) \ai\\
    &- g^2 \chi_b(\omega)\chi_a(\omega) \b,
\end{split}
\end{equation}

\begin{equation}
    \chi_b(\omega) = \left[ i(\omega_b - \omega) + \frac{\kappa_c}{2} \right]^{-1},
\end{equation}

\begin{equation}
    \chi_a(\omega) = \left[ i(\omega_\phi' - \omega) + \frac{\kappa_a}{2} \right]^{-1}.
\end{equation}

As we are in the small coupling regime, we ignore term with factor $g^2$. We then find an expression for the autocorrelation of $\b$ to be

\begin{equation}
\begin{split}
    \Sb =& \,\kappa_c\big{|}\chi_b(\omega)\big{|}^2\Sbi\\
    &+ g^2 \kappa_a \big{|}\chi_b(\omega)\big{|}^2 \big{|}\chi_a(\omega)\big{|}^2 \Sai.
\end{split}
\end{equation}

In the time domain, the correlation $\Sb$ has the form

\begin{equation}
    \mathcal{F}^{-1}\left[\Sb\right](\tau) = \langle\b^\dagger(t+\tau)\b(t)\rangle,
\end{equation}

which is the un-normalized $G^{(1)}$ correlation function of quantum optics. We now see that the correlation of the haloscope cavity is linked with the correlation of the input field of the ``axion" cavity.

\section{Photon Counting Experiments} \label{photonCounting}

This treatment is adapted from similar work in ref. \cite{mandel1995optical}.

\subsection{Single Photon Counting Probability}

We begin with a cavity in the vacuum state coupled to the axion field in some initial state 

\begin{equation}
    \rho(t_0) = \ri.
\end{equation}

We then use the integrated Von Neumann equation iterated to second order:

\begin{equation}
\begin{split}
      &\rho(\tf) =\\
      &\rho(\ti) + \frac{1}{i\hbar}\int_{\ti}^{\tf}dt\,\left[\Hi(t),\rho(\ti)\right]\\
      &+\frac{1}{(i\hbar)^2}\int_{\ti}^{\tf}dt'\int_{\ti}^{t'}dt''\,\bigg{[} \Hi(t'),\left[ \Hi(t'') , \rho(\ti) \right] \bigg{]}.
\end{split} 
\end{equation}

This is a good approximation for evolution over small timescales. We assume the state evolves for a time $\Delta t$. Constraints will be placed on this time later. We now want to know the probability that the system transitions to the state $\ket{1}\bra{1}\otimes\ket{\chi}\bra{\chi}$ for an arbitrary state of the axion field $\ket{\chi}$. The transition probability for the first two terms will be zero due to to orthogonality in the initial state $\rho(\ti)$. We then focus on the transition probability for the last term. The commutator in the last term gives us four total terms. Any terms where the density matrix is not surrounded by Hamiltonians on either side go to zero. We are then left with only two terms:

\begin{equation}
\begin{split}
    &\braket{1,\chi | \rho(\tf) | 1,\chi} =\\
    &\frac{1}{\hbar^2}\int_{\ti}^{\tf}dt'\int_{\ti}^{t'}dt''\,\braket{1,\chi | \Hi(t')\rho(\ti)\Hi(t'')|1,\chi}\\
    &+ \hc\,.
\end{split}
\end{equation}

We now expand this expression, using the interaction Hamiltonian rewritten using eq. \ref{eqn:fieldParts}

\begin{equation}
    \Hi = \frac{\sqrt{\hbar}}{c} g\left[\Ap(t)\bta e^{i\omega_bt} + \Am(t)\ba e^{-i\omega_bt}\right],
\end{equation}

 and noting that non-crossed terms will go to zero:

\begin{equation}
\begin{split}
    &\braket{1,\chi | \rho(\tf) | 1,\chi} =\\
    &\frac{g^2}{\hbar c^2}\int_{\ti}^{\tf}dt'\int_{\ti}^{t'}dt''\,\braket{\chi | \Ap(t')\rho_E\Am(t'') | \chi}\\
    &\times\big{|}\braket{0 | \b | 1}\big{|}^2 e^{-i\omega_b(t''-t')} + \hc.
\end{split}
\end{equation}

We now introduce cavity bandwidth effects. We consider that the cavity can absorb photons over a range of frequencies around resonance, and make the substitution

\begin{equation}
    \left|\braket{0 | \b | 1}\right|^2 e^{-i\omega_b(t''-t')} \to \int_{-\infty}^{\infty}d\omega\, H(\omega)e^{i\omega(t'-t'')},
\end{equation}

and define a function 

\begin{equation}
    k(\tau) = \frac{1}{2\pi}\int_{-\infty}^{\infty}d\omega\, H(\omega)e^{i\omega\tau}.
\end{equation}

Note that when $H(\omega) = 2\pi\delta(\omega-\omega_b)$, i.e., the cavity only absorbs photons on resonance, we recover the original expression.

We can now write the total expression as

\begin{equation}
\begin{split}
    &\braket{1,\chi | \rho(\tf) | 1,\chi} =\\
    &2\pi\frac{g^2}{\hbar c^2}\int_{\ti}^{\tf}dt'\int_{\ti}^{t'}dt''\,\braket{\chi | \Ap(t')\rho_E\Am(t'') | \chi}\\
    &\times k(t'-t'') + \hc\,.
\end{split}
\end{equation}

We now sum over an orthonormal basis for $\chi$ and recover the probability of the cavity transitioning to $\ket{1}$ regardless of the final axion state. We call this probability

\begin{equation}
\begin{split}
    &P(\ti)\Delta t =\\
    &2\pi\frac{g^2}{\hbar c^2}\int_{\ti}^{\tf}dt'\int_{\ti}^{t'}dt''\,k(t'-t'') \braket{\Am(t'')\Ap(t')}\\
    &+ \hc\,.
\end{split}
\end{equation}

We then make the Monochromatic approximation for the axion field. Consider the field is occupied around the central frequency $\omega_a$ at a spectral width $\Delta\omega_a$ and choose $\Delta t$ such that

\begin{equation}
    \frac{1}{\omega_a} \ll \Delta t \ll \frac{1}{\Delta \omega_a}.
\end{equation}

Using the fact that the quality factor of the field is approximately $10^6$, this is feasible and allows us to approximate

\begin{equation}
    \Ap(\tf) \approx \Ap(t_0)e^{-i\omega_a\Delta t}, 
\end{equation}

which allows us to simplify our equation to

\begin{equation}
\begin{split}
    &P(\ti)\Delta t = \\
    &2\pi\frac{g^2}{\hbar c^2}\braket{\Am(\ti)\Ap(\ti)}\\
    &\times \int_{\ti}^{\tf}dt'\int_{\ti}^{t'}dt''\,k(t'-t'')e^{-i\omega_a(t'-t'')}\\
    &+ \hc\,.
\end{split}
\end{equation}

After a change of variables and recombining terms, we find

\begin{equation}
\begin{split}
    &P(\ti)\Delta t =\\
    &2\pi\frac{g^2}{\hbar c^2}\braket{\Am(\ti)\Ap(\ti)}\int_{0}^{\Delta t}dt'\int_{-t'}^{t'}d\tau\,k(\tau)e^{-i\omega_a\tau}.
\end{split}
\end{equation}

We now impose another constraint on $\Delta t$,

\begin{equation}
    \Delta t \gg \frac{1}{\Delta\omega_b}.
\end{equation}

Because $\kappa(\tau)$ vanishes rapidly, we can say that

\begin{equation}
    \int_{-t'}^{t'}d\tau \approx \int_{-\infty}^{\infty}d\tau
\end{equation}

for almost all $t'\in (0, \Delta t)$. We simplify further and recover our final result

\begin{equation}
    P(\ti)\Delta t = 2\pi\frac{g^2}{\hbar c^2} H(\omega_a)\braket{\Am(\ti)\Ap(\ti)}\Delta t.
\end{equation}

We then see that the expectation value is equal to the un-normalized $G^{(1)}$ correlation function of the axion field evaluated at $\tau = 0$: 

\begin{equation}
    P(\ti)\Delta t = 2\pi\frac{g^2}{\hbar c^2} H(\omega_a)G^{(1)}(0)\Delta t.
\end{equation}

\subsection{Double Photon Counting Probability}

We begin assuming two separate cavities with the same resonant frequency, and we want to know the probability that both cavities absorb a photon during two disjoint time intervals $T_a = (\tia, \tfa)$ and $T_b = (\tib, \tfb)$. To calculate this, we will consider an interaction Hamiltonian that ``switches on" only during those time periods

\begin{equation}
\begin{split}
    &\Hi(t) =\\
    &\frac{\sqrt{\hbar}}{c} g\left[\Ap(t)\bta e^{i\omega_bt} + \Am(t)\ba e^{-i\omega_bt}\right]\ia(t)\\
    &+ \frac{\sqrt{\hbar}}{c} g\left[\Ap(t)\btb e^{i\omega_bt} + \Am(t)\bb e^{-i\omega_bt}\right]\ib(t),
\end{split}
\end{equation}

where $\ia$ and $\ib$ are the indicator functions of $T_a$ and $T_b$ respectively. We then use the Von Neumann equation iterated to fourth order to find the transition probability from the initial state 

\begin{equation}
    \rho(\ti) = \ket{0}\bra{0}_a \otimes \ket{0}\bra{0}_b \otimes \rho_E
\end{equation}

to the final state

\begin{equation}
    \rho(\ti + \Delta T) = \ket{1}\bra{1}_a \otimes \ket{1}\bra{1}_b \otimes \ket{\chi}\bra{\chi}
\end{equation}

assuming that $T_a,\,T_b \subset (\ti, \ti+\Delta T)$.

It is easy to see that all terms of the Von Neumann equation, except for the fourth-order term, will be zero. We are then left with

\begin{widetext}
\begin{align}
\begin{split}
    &\braket{1, 1, \chi | \rho(\TF) | 1, 1, \chi} = \\ 
    &\frac{1}{\hbar^4}\int_{\ti}^{\TF}d\tp\int_{\ti}^{\tp}d\tdp\int_{\ti}^{\tdp}d\ttp\int_{\ti}^{\ttp}d\tqp \braket{1, 1, \chi | \left[\Hi(\tp), \left[\Hi(\tdp), \left[\Hi(\ttp), \left[\Hi(\tqp), \rho(\ti)\right]\right]\right]\right] | 1, 1, \chi}.
\end{split}
\end{align}
\end{widetext}

We know that any terms where the initial density matrix is not surrounded by two Hamiltonians on each side will be zero and expand ignoring those terms:

\begin{widetext}
\begin{align}
\begin{split}
    &\braket{1, 1, \chi | \rho(\TF) | 1, 1, \chi} = \\
    &\frac{g^4}{\hbar^2 c^4}\int_{\ti}^{\TF}d\tp\int_{\ti}^{\tp}d\tdp\int_{\ti}^{\tdp}d\ttp\int_{\ti}^{\ttp}d\tqp\braket{\chi | \Ap(\tp)\Ap(\ttp)\rho_E\Am(\tqp)\Am(\tdp) | \chi}\left|\braket{0 | \ba | 1}\right|^2\left|\braket{0 | \bb | 1}\right|^2\\ 
    & \times \left( \ia(\tp)\ib(\ttp) + \ia(\ttp)\ib(\tp) \right)\left( \ia(\tqp)\ib(\tdp) + \ia(\tdp)\ib(\tqp)  \right)  e^{i\omega_a\tp} e^{i\omega_a\ttp} e^{-i\omega_a\tqp} e^{-i\omega_a\tdp} \\
    & + \frac{g^4}{\hbar^2 c^4}\int_{\ti}^{\TF}d\tp\int_{\ti}^{\tp}d\tdp\int_{\ti}^{\tdp}d\ttp\int_{\ti}^{\ttp}d\tqp\braket{\chi | \Ap(\tp)\Ap(\tqp)\rho_E\Am(\ttp)\Am(\tdp) | \chi}\left|\braket{0 | \ba | 1}\right|^2\left|\braket{0 | \bb | 1}\right|^2\\ 
    & \times \left( \ia(\tp)\ib(\tqp) + \ia(\tqp)\ib(\tp) \right)\left( \ia(\ttp)\ib(\tdp) + \ia(\tdp)\ib(\ttp)  \right)  e^{i\omega_a\tp} e^{i\omega_a\tqp} e^{-i\omega_a\ttp} e^{-i\omega_a\tdp} \\
    & + \frac{g^4}{\hbar^2 c^4}\int_{\ti}^{\TF}d\tp\int_{\ti}^{\tp}d\tdp\int_{\ti}^{\tdp}d\ttp\int_{\ti}^{\ttp}d\tqp\braket{\chi | \Ap(\tp)\Ap(\tdp)\rho_E\Am(\tqp)\Am(\ttp) | \chi}\left|\braket{0 | \ba | 1}\right|^2\left|\braket{0 | \bb | 1}\right|^2\\ 
    & \times \left( \ia(\tp)\ib(\tdp) + \ia(\tdp)\ib(\tp) \right)\left( \ia(\tqp)\ib(\ttp) + \ia(\ttp)\ib(\tqp)  \right)  e^{i\omega_a\tp} e^{i\omega_a\tdp} e^{-i\omega_a\tqp} e^{-i\omega_a\ttp} \\
    & + \hc\,.
\end{split}
\end{align}
\end{widetext}

We now solve this term by term.

\subsubsection{Term 1}

The indicator functions give us conditions on the order of $T_a$ and $T_b$ since $\tqp < \ttp < \tdp < \tp$. These conditions are as follows:

\begin{equation}
\begin{split}
    &\left[(\tp \in T_a \land \ttp \in T_b) \lor (\ttp \in T_a \land \tp \in T_b)\right] \\
    \land &\left[(\tqp \in T_a \land \tdp \in T_b) \lor (\tdp \in T_a \land \tqp \in T_b)\right].
\end{split}
\end{equation}

Without loss of generality, we will decide that $T_b$ comes before $T_a$. This implies

\begin{equation}
   \begin{split}
       &\tp,\,\tdp \in T_a, \\
       &\ttp,\,\tqp \in T_b.
   \end{split} 
\end{equation}

Then, by summing over all final states of the axion field, adding bandwidth effects (assuming identical cavities), and changing variables, we find term 1 is equal to

\begin{equation}
\begin{split}
    &(2\pi)^2\frac{g^4}{\hbar^2 c^4}\braket{\Am(\tib)\Am(\tia)\Ap(\tia)\Ap(\tib)}\\
    &\times \int_{0}^{\Delta t}d\tp\int_{0}^{\tp}d\tau'\int_{0}^{\Delta t}d\ttp\int_{0}^{\ttp}d\tau''\\
    &k(\tau')k(\tau'') e^{-i\omega_a\tau'}e^{-i\omega_a\tau''} + \hc\,.
\end{split}
\end{equation}

\subsubsection{Term 2}

The condition imposed by the indicator function here is

\begin{equation}
\begin{split}
    &\left[(\tp \in T_a \land \tqp \in T_b) \lor (\tqp \in T_a \land \tp \in T_b)\right] \\
    \land &\left[(\ttp \in T_a \land \tdp \in T_b) \lor (\tdp \in T_a \land \ttp \in T_b)\right].
\end{split}
\end{equation}

Again, assuming $T_b$ is before $T_a$ in time, we find

\begin{equation}
   \begin{split}
       &\tp,\,\tdp \in T_a, \\
       &\ttp,\,\tqp \in T_b.
   \end{split} 
\end{equation}

Using the same procedure as the last term, we find term 2 equal to

\begin{equation}
\begin{split}
    &(2\pi)^2\frac{g^4}{\hbar^2 c^4}\braket{\Am(\tib)\Am(\tia)\Ap(\tia)\Ap(\tib)}\\
    &\times\int_{0}^{\Delta t}d\tp\int_{0}^{\tp}d\tau'\int_{0}^{\Delta t}d\ttp\int_{-\ttp}^{0}d\tau''\\
    &k(\tau')k(\tau'') e^{-i\omega_a\tau'}e^{-i\omega_a\tau''}+ \hc\,.
\end{split}
\end{equation}

\subsubsection{Term 3}

The conditions given by the indicator functions are as follows:

\begin{equation}
    \begin{split}
        &\left[(\tp \in T_a \land \tdp \in T_b) \lor (\tdp \in T_a \land \tp \in T_b)\right] \\
        \land &\left[(\tqp \in T_a \land \ttp \in T_b) \lor (\ttp \in T_a \land \tqp \in T_b)\right].
    \end{split}
\end{equation}

These conditions are impossible to satisfy, as it would imply that either 

\begin{equation}
\begin{split}
    &\tp,\,\ttp \in T_a,\\
    &\tdp,\,\tqp \in T_b
\end{split}
\end{equation}

or 

\begin{equation}
\begin{split}
    &\tdp,\,\tqp \in T_a,\\
    &\tp,\,\tdp \in T_b.
\end{split}
\end{equation}

However, since the intervals $T_a$ and $T_b$ are assumed to be disjoint, and $\tqp < \ttp < \tdp < \tp$, this is impossible. Therefore, this term goes to zero no matter whether $T_a$ or $T_b$ is first.

\subsubsection{Final Expression}

By combining the above terms, we find

\begin{equation}
\begin{split}
    &P(\tia, \tib)(\Delta t)^2 =\\
    &(2\pi)^2\frac{g^4}{\hbar^2 c^4}\braket{\Am(\tib)\Am(\tia)\Ap(\tia)\Ap(\tib)}\\
    &\times\int_{0}^{\Delta t}d\tp\int_{0}^{\tp}d\tau'\int_{0}^{\Delta t}d\ttp\int_{-\ttp}^{\ttp}d\tau''\\
    &k(\tau')k(\tau'') e^{-i\omega_a\tau'}e^{-i\omega_a\tau''} + \hc\,.
\end{split}
\end{equation}

By imposing $\Delta t \gg 1/\Delta \omega_b$ and merging the conjugate, we find

\begin{equation}
\begin{split}
    &P(\tia, \tib)(\Delta t)^2 =\\
    &(2\pi)^2\frac{g^4}{\hbar^2 c^4} H(\omega_a)^2\braket{\Am(\tib)\Am(\tia)\Ap(\tia)\Ap(\tib)}(\Delta t)^2.
\end{split}
\end{equation}

Noting that the expectation value is the un-normalized second-order correlation function of the axion field, we write

\begin{equation}
    P(\tia, \tib)(\Delta t)^2 = (2\pi)^2\frac{g^4}{\hbar^2 c^4} H(\omega_a)^2G^{(2)}(t_a-t_b)(\Delta t)^2.
\end{equation}

This shows that we can measure the axion field's second order coherence by performing photon counting experiments on two haloscopes.

\end{document}